\documentstyle[12pt,aasms4,psfig]{article}
\evensidemargin=\oddsidemargin
\renewcommand{\baselinestretch}{2}
\def\ltsima{$\; \buildrel < \over \sim \;$}
\def\lsim{\lower.5ex\hbox{\ltsima}}
\def\gtsima{$\; \buildrel > \over \sim \;$}
\def\gsim{\lower.5ex\hbox{\gtsima}}

\newcommand{\be}{\begin{equation}}
\newcommand{\en}{\end{equation}}

\newcommand{\grb}{XRF\,060218}
\newcommand{\sn}{SN~2006aj}

\newcommand{\swift}{{\it Swift}}

\def\simlt{\mathrel{\hbox{\rlap{\hbox{\lower4pt\hbox{$\sim$}}}\hbox{$<$}}
}}
\def\simgt{\mathrel{\hbox{\rlap{\hbox{\lower4pt\hbox{$\sim$}}}\hbox{$>$}}
}}

\newcommand{\Msun}{\mbox{$M_\odot$}}
\newcommand{\Msunyr}{\mbox{$M_\odot$ yr$^{-1}$}}

\newcommand{\ebv}{$E_{B-V}$}

\newcommand{\kmsec}{\mbox{km s$^{-1}$}}

\begin{document}
\begin{center}

\vspace{1cm}

{\bf An optical supernova associated with X-ray flash 060218}

{E.~Pian$^{1,2}$, P.~A.~Mazzali$^{3,4,5,1,2}$, N.~Masetti$^6$,
P.~Ferrero$^7$, S.~Klose$^7$, E.~Palazzi$^6$, E.~Ramirez-Ruiz$^{8,9}$, 
S.~E.~Woosley$^9$, C.~Kouveliotou$^{10}$, J.~Deng$^{11,4,5,2}$, 
A.V.~Filippenko$^{12}$, R.~J.~Foley$^{12}$, J.~P.~U.~Fynbo$^{13}$, 
D.~A.~Kann$^7$, W.~Li$^{12}$, J.~Hjorth$^{13}$, K.~Nomoto$^{2,4,5}$, 
F.~Patat$^{14}$, D.~N.~Sauer$^{1,2}$, J.~Sollerman$^{13,15}$, 
P.~M.~Vreeswijk$^{16,17}$, E.~W.~Guenther$^7$, A.~Levan$^{2,18}$, 
P.~O'Brien$^{19}$, N.~R.~Tanvir$^{19}$, R.~A.~M.~J.~Wijers$^{20}$,  
C.~Dumas$^{17}$, O.~Hainaut$^{17}$, D.~S.~Wong$^{12}$, D.~Baade$^{14}$,  
L.~Wang$^{21,22}$, L.~Amati$^6$, E.~Cappellaro$^{23}$, 
A.~J.~Castro-Tirado$^{24}$, S.~Ellison$^{25}$, F.~Frontera$^{6,26}$, 
A.~S.~Fruchter$^{27}$, J.~Greiner$^{28}$, K.~Kawabata$^{29}$, 
C.~Ledoux$^{17}$, K.~Maeda$^{2,30}$, P.~M\o ller$^{14}$,
L.~Nicastro$^6$, E.~Rol$^{19}$, R.~Starling$^{20}$}

\end{center}

\noindent $^1$ Istituto Nazionale di Astrofisica, Trieste
Astronomical Observatory, via G.B. Tiepolo 11, I-34131 Trieste, Italy.

\noindent $^2$ Kavli Institute for Theoretical Physics,
University of California, Santa Barbara, California 93106-4030, USA.

\noindent $^3$ Max-Planck Institut f\"ur Astrophysik, Karl-Schwarzschild-Str. 
1, D-85748 Garching, Germany.

\noindent $^4$ Department of Astronomy, School of Science, 
University of Tokyo, Bunkyo-ku, Tokyo 113-0033, Japan.

\noindent $^5$ Research Center for the Early Universe, School of 
Science, University of Tokyo, Bunkyo-ku, Tokyo 113-0033, Japan.

\noindent $^6$ Istituto Nazionale di Astrofisica, IASF, Bologna, Via P. 
Gobetti 101, I-40129 Bologna, Italy.

\noindent $^7$ Th\"uringer Landessternwarte Tautenburg, Sternwarte 5, 
D-07778 Tautenburg, Germany.

\noindent $^8$ Institute for Advanced Study, Einstein Drive, Princeton, 
NJ 08540, USA.

\noindent $^9$ Department of Astronomy and Astrophysics, University 
of California, Santa Cruz, California 95064, USA.

\noindent $^{10}$ NASA/MSFC, NSSTC, VP62, 320 Sparkman Drive, 
Huntsville, Alabama 35805, USA.

\noindent $^{11}$ National Astronomical Observatories, Chinese Academy of 
Sciences, 20A Datun Road, Chaoyang District, Beijing 100012, China.

\noindent $^{12}$ Department of Astronomy, University of California, Berkeley, 
California 94720-3411, USA.

\noindent $^{13}$ Dark Cosmology Centre, Niels Bohr Institute, 
University of Copenhagen, Juliane Maries Vej 30, DK-2100 Copenhagen \O , 
Denmark.

\noindent $^{14}$ European Southern Observatory, Karl-Schwarzschild-Str. 2, 
D-85748 Garching bei M\"unchen, Germany.

\noindent $^{15}$ Stockholm Observatory, Department of Astronomy, AlbaNova, 106 91 
Stockholm, Sweden

\noindent $^{16}$ Departamento de Astronom\'ia, Universidad de 
Chile, Casilla 36-D, Santiago, Chile.

\noindent $^{17}$ European Southern Observatory, Alonso de C\'ordova 
3107, Casilla 19001, Santiago 19, Chile.

\noindent $^{18}$ Centre for Astrophysics Research, University of 
Hertfordshire, College Lane, Hatfield AL10 9AB, United Kingdom.

\noindent $^{19}$ X-Ray and Observational Astronomy Group, 
Department of Physics \& Astronomy, University of Leicester,
Leicester LE1 7RH, United Kingdom.

\noindent $^{20}$ Astronomical Institute ``Anton Pannekoek", University of 
Amsterdam, Kruislaan 403, 1098 SJ Amsterdam, The Netherlands.

\noindent $^{21}$ Lawrence Berkeley National Laboratory, 1
Cyclotron Road, Berkeley, CA 94720, USA.

\noindent $^{22}$ Purple Mountain Observatory, Chinese Academy of
Sciences, 2 Beijing Xi Lu, Nanjing, Jiangsu 210008, China.

\noindent $^{23}$ Istituto Nazionale di Astrofisica, Padova Astronomical 
Observatory, Vicolo dell'Osservatorio 5, I-35122 Padova, Italy.

\noindent $^{24}$ Instituto de Astrofisica de Andalucia (IAA-CSIC), 
Apartado de Correos 3004, 18080 Granada, Spain.

\noindent $^{25}$ Department of Physics and Astronomy, University of 
Victoria, 3800 Finnerty Road, Victoria, BC, Canada V8P 1A1.

\noindent $^{26}$ Department of Physics, University of Ferrara, Polo Scientifico 
e Tecnologico, Edificio C, via Saragat 1, I-44100 Ferrara, Italy

\noindent $^{27}$ Space Telescope Science Institute, 3700 San Martin 
Drive, Baltimore, Maryland 21218, USA.

\noindent $^{28}$ Max-Planck-Institut f\"ur extraterrestrische 
Physik, Giessenbachstrasse, D-85741 Garching, Germany.

\noindent $^{29}$ Hiroshima Astrophysical Science Center, 
Hiroshima University, Hiroshima 739-8526, Japan.

\noindent $^{30}$ Department of Earth Science and Astronomy, 
College of Arts and Sciences, University of Tokyo, Komaba 3-8-1, 
Meguro-ku, Tokyo 153-8902, Japan.

\newpage



{\bf Long-duration gamma-ray bursts (GRBs) are associated with Type Ic
supernovae$^1$  that are  more luminous  than the  average$^{2-5}$ and
eject material at very high velocities.  Less luminous supernovae were
not  hitherto  known  to   be  associated  with  GRBs,  and  therefore
GRB-supernovae were  thought to be  rare events$^6$. The  detection of
X-ray flashes (XRFs) -- analogues of GRBs, but with lower luminosities
and  fewer gamma-rays --  raised the  issue of  whether they  are also
associated  with supernovae  and whether  they are  intrinsically weak
events or  typical GRBs  viewed off-axis$^7$.  Here  we report  on the
optical  discovery  and follow-up  of  the  Type  Ic supernova  2006aj
associated with \grb.  SN~2006aj  was intrinsically less luminous than
the  GRB-supernovae,  but  more  luminous  than  many  supernovae  not
accompanied by a GRB.  The  ejecta velocities derived from our spectra
are intermediate  between these two  groups, which is  consistent with
the  weakness of  both  the  GRB output$^8$  and  the supernova  radio
flux$^9$.     Our    data,    combined    with   radio    and    X-ray
observations$^{8-10}$, suggest that \grb\ is an intrinsically weak and
soft  event,  rather than  a  classical  GRB  observed off-axis.   The
discovery  of  \sn\  and   its  properties  extend  the  GRB-supernova
connection  to  XRFs and  to  fainter  supernovae,  implying a  common
origin.   Events such as \grb\   probably  dominate  in   number  over
GRB-supernovae.}


The Burst Alert Telescope (BAT) onboard \swift\ detected \grb\ on 2006
February  18, 03:34:30  UT$^8$. Its spectrum  peaked  near 5\,keV,
placing  the  burst  in  the   XRF  subgroup  of  GRBs.   The  optical
counterpart of the burst was detected $\sim$200~s later by the \swift\
Ultraviolet/Optical  Telescope,  and   was  subsequently  observed  by
ground-based telescopes$^{11}$.   The closeness of the event$^{12}$ made
\grb\ an ideal   candidate  for  spectroscopic   observations  of   a  possible
associated supernova.

We observed \grb\ with  the European Southern Observatory's (ESO) 8.2m
Very  Large Telescope (VLT)  and the  University of  California's Lick
Observatory  Shane 3~m  telescope  (Lick) starting  21 February  2006.
Table~1  in   the  Supplementary  Information shows  the  log   of  the
observations.  Spectroscopy  was performed nearly  daily for seventeen
days (see  Figure~1 in the Supplementary  Information).  Broad absorption
lines detected in  our first spectrum resembled  those  of broad-lined 
Type Ic supernovae, thus providing the first definite case of a supernova
associated  with an
XRF$^{13}$.   This  is  the  earliest  spectroscopy  of  a
GRB-supernova,  and in  fact one  of the  earliest for  any supernova.
Based on its  early decline, we estimate that  the contribution of the
fading afterglow of \grb\ to the supernova emission is not significant
at the epoch of our first spectrum$^{11,12}$.

The high-dispersion spectrum taken with the VLT Ultraviolet and Visual
Echelle  Spectrograph  (UVES)  near  the epoch  of  supernova  maximum
exhibits  several  narrow emission  and  absorption  lines.  From  the
former  we  obtained  an   accurate  measurement  of  the  host-galaxy
redshift,  $z   =  0.03342  \pm   0.00002$  (heliocentric  corrected),
corresponding to a distance of  $\sim140$ Mpc (using a Hubble constant
of  $H_0  = 73$  km~s$^{-1}$~Mpc$^{-1}$, $\Omega_\Lambda=0.72$,  and
$\Omega_m=0.28$).   We  constrained the  total  extinction toward  the
supernova  from  the  equivalent  widths of  the  interstellar  Na~I~D
absorption  lines$^{14}$  to  be  \ebv  = $0.13  \pm  0.02$  mag
(P.A.M.,  manuscript  in preparation).   The extinction  is  mainly  
due to  our
Galaxy,  and its value is consistent  with  that  derived  using infrared  dust
maps$^{15}$.  We  used this value  to correct the light  curve of
\sn\ (Figure 1).

It  is interesting to  compare the  properties of  \sn\ with  those of
other Type Ic supernovae.    The  three   well-observed,   low-redshift
GRB-supernovae (SN~1998bw,  2003dh and 2003lw) are  striking for their
similarities.   They are  $\sim$5-6 times  more luminous  and $\sim$30
times more  energetic than typical type-Ic supernovae$^{16}$. The 
peak luminosities
and the  kinetic energies  of  the  GRB-supernovae differ  by no more 
than 30\%.  At  maximum light, \sn\ is dimmer  than these supernovae
by about  a factor  of 2,  but it  is still a  factor of  2 to  3 more
luminous than other broad-lined Ic supernovae not associated with GRBs
and normal (i.e, narrow-lined) supernovae Ic (Figure 1).

Normal type Ic supernovae rise to a peak in approximately 10-12 days and have
photospheric expansion  velocities of $\sim$10000  \kmsec\ at $\sim$10
days. Previously known GRB-supernovae showed a longer risetime (14--15
days) and had, at an epoch of $\sim$10 days, velocities of $\sim$25000
\kmsec\  (see  Figures   1  and  2).   If  \grb\   and  \sn\  occurred
simultaneously,  \sn\  rose  as  fast  as normal  supernovae  Ic,  and
declined also comparatively fast.   At the same time, the photospheric
expansion  velocity derived  from spectral  modelling  is intermediate
between  the GRB-supernovae  and other  supernovae Ic,  broad-lined or
narrow-lined,  that   were  not  associated  with   GRBs  (Figure  2).
Asymmetry  in   the  supernova  explosion  may   modify  the  observed
luminosity  with  respect  to  the  intrinsic one,  depending  on  the
orientation  of  the  symmetry  axis, by no more than 25\%  (ref. 17).

We  conclude  that \sn\  is  intrinsically  dimmer  than the  other  3
GRB-supernovae.  In  addition, it is associated with  the softest (but
not   the  weakest)   of  the   four  local   events   connected  with
supernovae$^8$,  and it has  mildly relativistic  ejecta$^{8,9}$, thus
appearing  to be  an  intermediate object  between GRB-supernovae  and
other type Ic supernovae, both broad-lined and narrow-lined, not accompanied
by a GRB.

All together,  these  facts point  to  a  substantial diversity  between
supernovae associated  with GRBs and supernovae  associated with XRFs.
This diversity  may be related to  the masses of  the exploding stars.
In a companion paper, the parameters of the explosion are derived from
models  of the  supernova  optical  light curves  and  spectra, and  a
relatively low initial mass, 20  \Msun, is proposed, evolving to a 3.3
\Msun\ CO star$^{18}$.   This mass is smaller than those estimated for
the typical GRB-supernovae$^{19}$.

GRBs and GRB-supernovae are aspherical  sources. If  XRF060218 was a normal 
GRB viewed off-axis, the observed soft
flux was emitted at large angles with respect to its jet axis.
If the  associated  \sn\  is
aspherical, then  it is also probably seen  off-axis.  Alternatively,
\grb\ may have  been intrinsically soft, whether it  was an aspherical
explosion  viewed on  axis  or a  spherical  event. Various  independent
arguments,  like  the   chromatic  behaviour  of  the  multiwavelength
counterpart of \grb$^8$, the  absence of a late radio rebrightening$^9$
and  the compliance of  \grb\ with  the empirical  correlation between
peak   energy   and  isotropic   energy$^{10}$,   favour  the   latter
possibility.

Together with the observation of other underluminous, relatively nearby XRFs
and GRBs -- GRB~980425 (ref. 2), XRF~030723 (refs. 20,21), XRF~020903 (ref. 22),
and GRB~031203 (refs. 23,24), some definitely and some probably associated with
supernovae -- the properties of \grb\ suggest the existence of a population
of events less luminous than ``classical'' GRBs, but possibly much more
numerous and with lower radio luminosities$^9$. Indeed, 
these events may be the most abundant form of X- or
gamma-ray explosive transient in the Universe, but instrumental limits
allow us to detect them only locally, so that several intrinsically
sub-luminous bursts may remain undetected. The fraction of
supernovae that are associated with GRBs or XRFs may be higher than
currently thought. 

By including this underluminous population and assuming no correction for
possible collimation, which may vary from object to object, we obtain a local
GRB rate of $110^{+180}_{-20}\;{\rm Gpc^{-3} yr^{-1}}$, compared to $1\;{\rm
Gpc^{-3} yr^{-1}}$ estimated from the cosmological events only (see
Supplementary Information for details).  In particular, for the detection
threshold of {\it Swift}, we expect a few bursts per year within $z = 0.1$
and with luminosities as low as that of GRB~980425. The low-energy GRB
population could be part of a continuum of explosion phenomena that mark the
collapse of a stellar core, with normal supernovae at one end and classical
GRBs at the other.

\noindent
{\bf Supplementary Information} is linked to the online version of the 
paper at www.nature.com/nature.

\bigskip

\noindent
{\bf Acknowledgements}  This work  is based on  data collected  by the
GRACE  consortium   with  ESO  Paranal  telescopes.    The  ESO  staff
astronomers  at  Paranal   are  acknowledged  for  their  professional
assistance.  We are grateful to S.~R.~Kulkarni, M.~Modjaz, A.~Rau, and
S.~Savaglio for helpful interactions  and to R.~Wilman for allowing us
to  implement our  Target-of-Opportunity program  with VLT  during his
scheduled  observing  time.  We  thank S.~Barthelmy  for  providing
information about the Swift/BAT performance.  This work has benefitted
from  collaboration  within  the  EU  FP5  Research  Training  Network
``Gamma-Ray Bursts:  an Enigma and  a Tool''.  IRAF is  distributed by
the National  Optical Astronomy  Observatories, which are  operated by
the Association of Universities  for Research in Astronomy, Inc, under
contract to  the National Science Foundation (NSF).  A.V.F.'s group at
UC  Berkeley  is supported  by  NSF  and  by the  TABASGO  Foundation.

\bigskip

\noindent
{\bf  Author  Contributions}  E.Pian,  N.M.,  P.F.,  S.K.,  E.Palazzi,
A.V.F., R.J.F., W.L., F.P.,  P.M.V., E.W.G., C.D., O.H., D.S.W., D.B.,
L.W., S.E., C.L. organized the observations and were responsible for data
acquisition, reduction and analysis; P.A.M., E.R.-R.,
S.E.W., J.D., K.N., D.N.S., K.M. contributed to the interpretation and
discussion  of the data;  J.P.U.F, D.A.K.,  J.H., J.S.,  A.L., P.O.B.,
L.A.,  E.C.,   A.J.C.-T.,  F.F.,  A.S.F.,  J.G.,   K.K.,  P.M.,  L.N.,
E.R.  provided expertise on specific aspects  of  the data
presentation  and discussion; E.Pian,  P.A.M., E.R.-R.,  S.E.W., C.K.,
K.N., N.R.T., R.A.M.J.W., E.C., R.S.  have written the manuscript.

\bigskip

\noindent
{\bf Author  Information} The  authors declare no  competing financial
interest.  Correspondence   and  requests  for   materials  should  be
addressed to E.P. (pian@oats.inaf.it).

\clearpage

\renewcommand{\baselinestretch}{1}

\begin{figure}
\centerline{\psfig{file=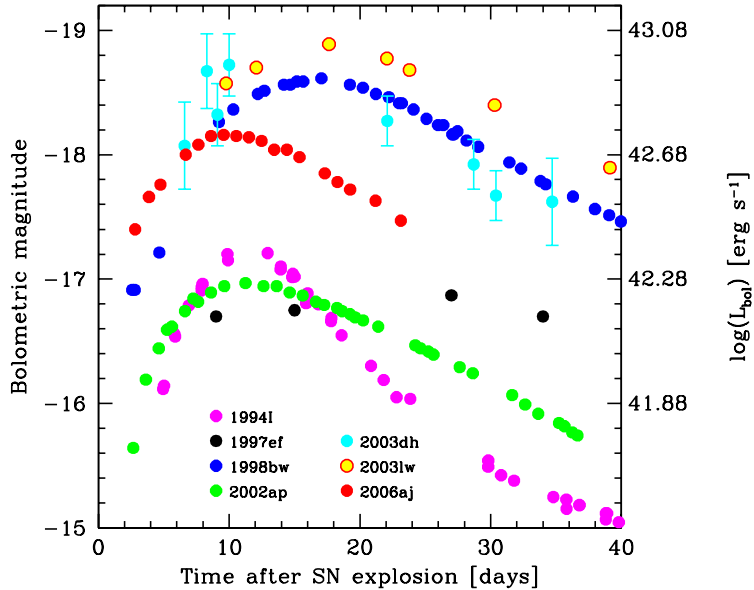,width=12.cm,angle=0}}
\caption[]  {{\bf  Bolometric light  curves  of  Type Ic  Supernovae.}
We report, as a function of time, the luminosity and corresponding
absolute magnitude of the four  
spectroscopically  identified
supernovae associated with GRB/XRFs, namely SN~1998bw (GRB~980425, $z =
0.0085$),  2003dh (GRB~030329, $z  = 0.168$),  2003lw (GRB~031203,  $z =
0.1055$),  and  2006aj  (\grb, $z  = 0.03342$);   
of  2  broad-lined  supernovae  (not
accompanied  by  a  GRB),  1997ef  and  2002ap;  and  of  the  normal,
intensively  monitored SN~1994I. All  represented supernovae  are Type
Ic.   The  light curves,  reported  in  their  rest frame,  have  been
constructed in the 3000--24000  \AA$\;$ range, taking into account the
Galactic     and,    where     appropriate,     the    host     galaxy
extinction$^{16,25-28}$.    For  \sn,  we
used the optical  light curves obtained during our  monitoring and the
near-infrared data  reported by ref. 29,  and a total
extinction value of  $E_{B-V} = 0.13$ mag (see  text).  We adopted the
extinction curve of ref. 30 with $R_V = 3.1$.  The
galaxy contribution  has also been subtracted  where significant.  The
initial time has  been assumed to coincide with  the XRF detection time,
2006 Feb 18.149  UT.  The systematic errors (about  0.2 mag) have been
omitted,  for  clarity.  The  shape  of the  light  curve  of \sn\  is
strikingly  similar   to  that  of   SN~2002ap,  as  are   indeed  the
spectra$^{18}$.
\label{fig:bollc}}
\end{figure}  


\begin{figure}
\centerline{\psfig{file=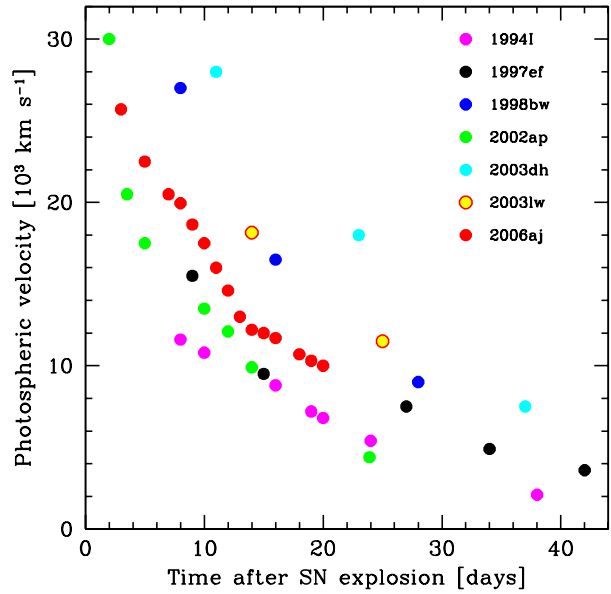,width=12.cm,angle=0}}
\caption[]  {{\bf   Photospheric  expansion  velocities   of  Type  Ic
Supernovae.}   The time profiles  of the  expansion velocities  of the
same  seven  supernovae  represented  in  Fig.~1  are  reported.   The
velocities have been  determined through models of the  spectra at the
various epochs$^{16,18,25,26}$.
\label{fig:vexp}}
\end{figure}  


\clearpage



\noindent
{\bf Supplementary Information
for ``An optical supernova associated with X-ray flash 060218''}

\vspace{1cm}


{\bf 1~~~Supplementary Table}

\bigskip

\noindent
In this Table we present the log of the observations, with details on the
acquisition, and part of the photometric results. 
Low-resolution spectra of \sn\ have been obtained with the ESO
VLT  UT1   and  UT2  telescopes,  equipped  with   the  FOcal  Reducer
Spectrographs (FORS1  and FORS2), and with the  Lick telescope, equipped
with the Kast  Dual-Beam Spectrograph (KDBS) with the  D55 dichroic. A
high-resolution spectrum of  the source was acquired with  the VLT UT2
equipped with UVES.  Photometry in the $BVRI$ bands has been performed
simultaneously with  each spectrum  with the VLT  and with  the 0.76~m
Katzman  Automatic   Imaging  Telescope  (limited   to  the  Lick+KDBS
observation),  except  on March  8,  when  only  $BVR$ photometry  was
acquired$^{31}$.  Spectroscopic monitoring was discontinued on 10 March 2006,
due to  Sun elevation constraints.  However,  photometry was performed
for a  few more  days with  the VLT.  In Column~1 we report the
observation date,  in Column~2 the telescope and instrument used, in 
Column~3 the observing setup, in Column~4 the exposure times of the 
spectra, in Column~5 the seeing during the spectrum acquisition, in Column~6
the  magnitudes not  corrected  for  the  Galactic  and  intrinsic
extinction, nor for the  host-galaxy flux contribution, and in Column~7
the same magnitudes reported in Column~6, but corrected for  
the host-galaxy  flux.  The
associated errors are 1$\sigma$ uncertainties.

\clearpage

\begin{table}
\begin{center}
\setlength{\tabcolsep}{4pt}
\begin{tabular}{ccccccc}
\multicolumn{7}{c}{\bf Supplementary Table 1: 
Summary of observations of \sn.}\\
\hline
\hline
Date  & Telescope+ & Setup & Integr. & Seeing  & $V$ & $V_{sub}$ \\
(2006 UT) & Instrument & & Time  (s) & (arcsec) & magnitude & magnitude \\
\hline
Feb 21.041 & UT1+FORS2 & 300V+GG435            & 1800 & 1.70 & $18.17 \pm 0.03$ & $18.36 \pm 0.04$ \\
Feb 22.159 & Lick+KDBS & D55+600/4310          & 6000 & 2.02 & $17.92 \pm 0.08$ & $18.06 \pm 0.09$ \\
           &           &             +300/7500 &      &      &                  & \\
Feb 23.026 & UT1+FORS2 & 300V                  & 1800 & 1.68 & $17.80 \pm 0.03$ & $17.93 \pm 0.03$ \\
Feb 25.023 & UT1+FORS2 & 300V                  & 1800 & 1.13 & $17.58 \pm 0.03$ & $17.68 \pm 0.03$ \\
Feb 26.016 & UT1+FORS2 & 300V                  & 1800 & 1.08 & $17.51 \pm 0.03$ & $17.61 \pm 0.03$ \\
Feb 27.023 & UT2+FORS1 & 300V                  & 1800 & 1.77 & $17.46 \pm 0.03$ & $17.55 \pm 0.03$ \\
Feb 28.025 & UT2+FORS1 & 300V                  & 2593 & 1.14 & $17.45 \pm 0.03$ & $17.54 \pm 0.03$ \\
Mar 01.009 & UT1+FORS2 & 300V+GG435            & 1800 & 1.14 & $17.45 \pm 0.03$ & $17.54 \pm 0.03$ \\
Mar 02.007 & UT1+FORS2 & 300V                  & 1800 & 1.63 & $17.47 \pm 0.03$ & $17.56 \pm 0.03$ \\
Mar 03.010 & UT1+FORS2 & 300V                  & 1800 & 1.12 & $17.51 \pm 0.03$ & $17.61 \pm 0.03$ \\
Mar 04.009 & UT1+FORS2 & 300V                  & 1800 & 1.26 & $17.56 \pm 0.03$ & $17.66 \pm 0.03$ \\
Mar 04.021 & UT2+UVES  & Dic. 1/390B           & 2100 & 1.26 &                  &                  \\
           &           &            +564R      &      &      &                  &                  \\
Mar 05.027 & UT2+FORS1 & 300V                  & 1350 & 0.83 & $17.60 \pm 0.03$ & $17.71 \pm 0.03$ \\
Mar 06.014 & UT1+FORS2 & 300V                  & 1800 & 1.70 & $17.68 \pm 0.03$ & $17.79 \pm 0.03$ \\
Mar 08.007 & UT2+FORS1 & 300V                  & 1800 & 1.89 & $17.86 \pm 0.03$ & $18.00 \pm 0.03$ \\
Mar 09.013 & UT2+FORS1 & 300V                  & 1800 & 0.95 & $17.92 \pm 0.03$ & $18.06 \pm 0.03$ \\
Mar 10.013 & UT2+FORS1 & 300V                  & 1560 & 1.40 & $18.01 \pm 0.03$ & $18.17 \pm 0.03$ \\
\hline
\end{tabular}
\end{center}
\end{table}

\clearpage


{\bf 2~~~Supplementary Figure and Legend}

\bigskip

\noindent
In this Figure we report the VLT FORS1/2 and Lick
spectra of  \sn\ taken at a  resolution of 3.4  \AA/pixel (FORS1), 2.6
\AA/pixel (FORS2),  and 1.9/4.6 \AA/pixel  (Lick, blue and  red sides,
respectively),  reduced  to rest  frame  and  scaled  up by  arbitrary
factors  for  clarity.  For  each  spectrum  the  time  elapsed  since
\grb\ explosion (Feb 18.149,  2006) is  indicated, in  days.  The
spectroscopic  and  photometric  data   (see  Supplementary Table  1)  were  reduced
following standard procedures within the IRAF and MIDAS data reduction
packages, respectively. IDL routines  were also used for the reduction
of the Lick spectrum.  Telluric absorption features have been removed.
To  account  for slit  losses,  the  spectra  were normalized  to  the
simultaneous $V$-band photometry: each spectrum was convolved with the
response function of the Bessell $V$ filter and the scaling factor was
determined  by  comparison   with  the  $V$-band  measured  magnitude.
Finally, the contribution of  the host-galaxy continuum was subtracted
from the spectra and  photometry, by linearly interpolating its 
fluxes$^{32-34}$.   No correction for interstellar reddening  was applied to
the spectra.   Owing to  the large airmass  at which  the observations
were performed, the relative flux calibration of the spectra shortward
of $\sim$4500~\AA\  is not  completely reliable.  The  low-contrast features
visible in some  VLT spectra longward of $\sim$9000~\AA\  are also not
meaningful.   The spectra  show broad  absorption lines  indicative of
high-velocity ejecta,  comparable to those present  in other energetic
Type  Ic supernovae$^{35}$, although  not  as high  as in  typical
GRB-supernovae$^{36-40}$.  
Superimposed  on the spectra are  emission lines from
the  host  galaxy.   Using  the   H$\alpha$  and  [O  {\sc  II}]  line
luminosities   we   derive$^{41}$  a   star-formation  rate   of
$\sim$0.06~\Msunyr.

\clearpage

\begin{figure}
\caption[]  {{\bf  Supplementary Figure 1: 
Spectra of SN~2006aj acquired with the VLT and Lick telescopes.}   
\label{fig:spectra}}
\centerline{\psfig{file=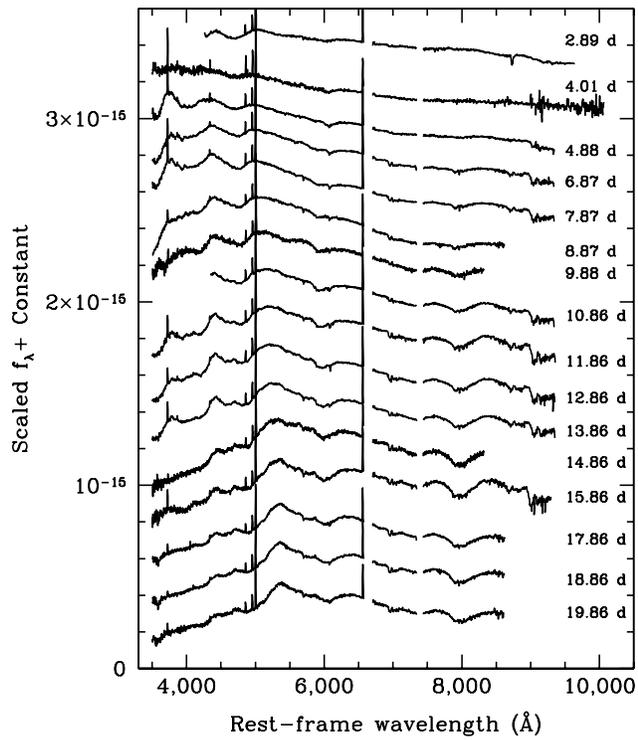,width=12.cm,angle=0}}
\end{figure}  

\clearpage

{\bf 3~~~Supplementary Methods}

\bigskip

\noindent
{\bf The rate of low luminosity GRBs and XRFs.}

\bigskip

\noindent
If the low-redshift GRBs are really typical of the global GRB population,
then their discovery within the current time and sky coverage must be
consistent with the local GRB explosion rate as deduced from the very
large BATSE GRB sample. In this section, we study under which
conditions low-redshift events can be derived from a luminosity function
that is consistent with the $\log N-\log S$ relationship for
``classical'' cosmological bursts.

All local rate estimates made prior to the discovery of GRB~031203 were
derived under the hypothesis that classical bursts greatly exceed a
minimum luminosity, $L_{\rm min}$, of about $5\times 10^{49}$
erg~s$^{-1}$. It was not until the discovery of GRB~031203 that it became clear
that the three nearby bursts, 980425, 030329 and 031203, were not
consistent with a population of bursts with luminosities greatly
exceeding that of GRB~980425 (refs. 42,43). The discovery by {\em
Swift} of the underluminous \grb\ slightly after one year of
operation gives further credence to this hypothesis. A unified picture
can therefore only be achieved by extending down the luminosity
function.

The luminosity function used here is based on an extension down to the
lowest luminosities consistent with the BATSE cumulative distribution
of the number of GRBs as a function of their fluence ($\log N - \log S$),
and at the same time gives the correct number of low-redshift events
as collected by BATSE, {\it HETE-II} and {\it Swift}.

The luminosity function is characterized by a smoothed broken
power-law,
\begin{equation}
\Phi(L)=\Phi_0\left[\left({L \over L_b}\right)^{\alpha} +\left({L \over
L_b}\right)^{\beta}\right]^{-1},
\end{equation}
where $L$ is the isotropic equivalent luminosity and does not take
into account the effects of collimation. The number of bursts with a
peak flux $>P$ is then given by:
\begin{equation}
N(>P)=\int_{L_{\rm min}}^{L_{\rm max}}\Phi(L)d\log L
\int_0^{z_{\rm max}(L,P)} \frac{R_{\rm GRB}(z)}{1+z}
\frac{dV(z)}{dz}dz
\end{equation}
where $dV(z)/dz$ is the comoving volume element, which in a flat
$\Lambda$CDM universe, is given by
\begin{equation}
{dV \over dz}={c \over H_0}{D^2_L \over (1+z)^2}{1 \over
(\Omega_M(1+z)^3+\Omega_\Lambda)^{1/2}}.
\end{equation}

That such an analysis will be possible follows from the
currently-favored idea that GRBs trace the star formation history of
the Universe: $R_{\rm GRB}(z)=R_{\rm SFR}(z)$. An analytic formula for
the cosmic star formation rate per unit comoving volume is adopted
here, as given in ref. 44.

The shape of the luminosity function is constrained here by two
different methods. First, similarly to ref. 42 we fit the model to the
peak flux distribution observed by BATSE (all 2204 bursts from the
GUSBAD catalog) by assuming an average {\it rest frame} GRB spectrum
with a peak energy of $200\pm 50$ keV and a low (high) energy photon
index of $-1\pm 0.5$ ($-2\pm0.5$). The model predictions are then
compared to the redshift and luminosities of GRBs detected by BATSE,
{\it HETE-II} and {\it Swift}, where the sensitivity curves of all
three instruments have been used$^{45}$. The individual
constraints are subsequently combined to derive the luminosity
function's best-fit parameters.

The major uncertainty in the above method concerns $L_{\rm min}$,
which we fix to be equal to the luminosity of GRB~980425. 
By  using $L_{\rm  max}=6 \times  10^{52}$ erg~s$^{-1}$,  $L_{\rm b}=9
\times 10^{50}$ erg~s$^{-1}$,  $\alpha=0.3$, $\beta=0.95$ in Equations
(1) and (2)  above we obtain  the best statistical description  of the
data  and  a  local   GRB  rate  of  $110^{+180}_{-20}\;{\rm  Gpc^{-3}
yr^{-1}}$.

The local rate of events that
give rise to GRBs is therefore at least one hundred times the rate
estimated from the cosmological events only (i.e. those observed by
BATSE). Interestingly, we find that a single power-law description for
the luminosity function is rejected with fairly high confidence and
that an intrinsic break in the luminosity function is indeed required.

Obviously, the above calculation is only sketchy and should be taken
as an order of magnitude estimate at present, as the observed redshift
distributions are likely to be plagued by severe selection effects. It
should, however, improve as more bursts with known redshifts are
detected. This estimate is nonetheless consistent with the current
rate of low-redshift events and is broadly in agreement with
conclusions from earlier statistical studies$^{42}$.

\clearpage

{\bf 4~~~Supplementary Notes}

\bigskip

\noindent
We report here the references for the previous Supplementary Section 3.

\end{document}